\documentclass[a4paper,10pt,twoside]{cpc-hepnp}

\usepackage{multicol}
\usepackage{graphicx}
\usepackage{booktabs}
\usepackage{amssymb,bm,mathrsfs,bbm,amscd}
\usepackage[tbtags]{amsmath}
\usepackage{lastpage}

\begin{document}

\fancyhead[c]{\small “Submitted to ‘Chinese Physics C'} \fancyfoot[C]{\small 010201-\thepage}

\footnotetext[0]{Received 14 March 2009}

\title{Design of a 325MHz SC Spoke040 cavity at IHEP 
}

\author{%
      PENG Ying-Hua(彭应华)$^{1,2;1)}$\email{pengyh@mail.ihep.ac.cn}%
\quad ZHANG Xin-Yin(张新颖)$^{1,2;2)}$\email{zhangxy@mail.ihep.ac.cn}%
\quad SHA Peng（沙鹏）$^{2}$\\
\quad PAN Wei-Min（潘卫民）$^{2}$
\quad LI Han(李菡)$^{1,2}$\\
}
\maketitle

\address{%
$^1$  Graduate University of Chinese Academy of Science, Beijing 100049, China\\
$^2$  Institute of High Energy Physics, Chinese Academy of Sciences, Beijing 100049, China\\
}

\begin{abstract}
The 325MHz, β=0.40 superconducting single spoke cavity (Spoke040) was one of the most challenges for the China-ADS (Accelerator Driven System) project. The design was finished, and the fabrication was in progressing. In this paper, we studied the main radio frequency (RF) and mechanical parameters, and compared the arc-shaped and plate-shaped end group structures..
\end{abstract}




\begin{multicols}{2}

\section{Introduction}

The successful horizontal test (H.T.) of the first 325MHz, β=0.12 superconducting single spoke cavity in IHEP in September 2013 have made a milestone in the R\&S of spoke cavities in China\cite{lab1}. Spoke cavities operate in the fundamental TEM-mode, tested to date are designed for particle velocities $0.15<\beta<0.75$\cite{lab2} A serious of 325MHz, β=0.40 superconducting single spoke cavities had been designed to accelerate a 10mA proton beam from 34MeV to 178MeV $(0.26<\beta<0.54)$ , next to Spoke021 cavities and then superconducting elliptical cavities in the layout of China-ADS linac\cite{lab2}.

With the experience of successful design, fabrication, processing and testing of the first two 325MHz,Spoke012 in IHEP, we were more confidence in the study of SC spoke cavities. There are also many challenges in the development of Spoke040. Its’ mainly size was much larger than that of Spoke012 and Spoke021, which need more concern on the mechanical properties and the facture of
the mold equipments. It would be the first time to study $\beta=0.40$ single SC spoke cavity in the world.

In this paper, RF design parameters and mechanical properties were respectively studied by CST microwave Studio and Ansys workbench. A new design for the end group would also been proposed.

\section{RF design}

The purpose of RF design is to get lower heat load and a higher accelerating gradient, that is get a higher $R/Q_0$  and lower peak surface fields ($Epeak/Eacc$ and $Bpeak/Eacc$)\cite{lab3}.  Generally $Epeak/Eacc$ should be slightly less than 3\cite{lab4}.
Before doing the electromagnetic optimizations of the Spoke040, several parameters have been chosen in order to match China-ADS general requirements:325MHz frequency;
Proton beam energy area: 34MeV-178MeV$(0.26<\beta<0.54)$;
$\beta_g=0.40$;
Total length $<614mm$ (including the liquid helium bath and tuner);
Vcmax=2.86MV, come from $Eacc= Vc/ \beta\lambda =7.7MV;
Epeak<32.5MV/m, Bpeak<65mT$.

And we summarized the principles for designing Spoke040 cavity:
Minimize the Epeak and Bpeak to match the physics requirements;
Minimize the length and radius of the cavity to make it more tight to use the already existing V.T. (vertical test) facilities and the end flanges;
Simple the structure, make it easier to fabrication and progress;
R/Q  and G should be as large as possible to achieve
promising $Q_0$,decrease the $P_c$.

\subsection{The sensitive analyze}

\begin{center}
\includegraphics[width=4.0cm]{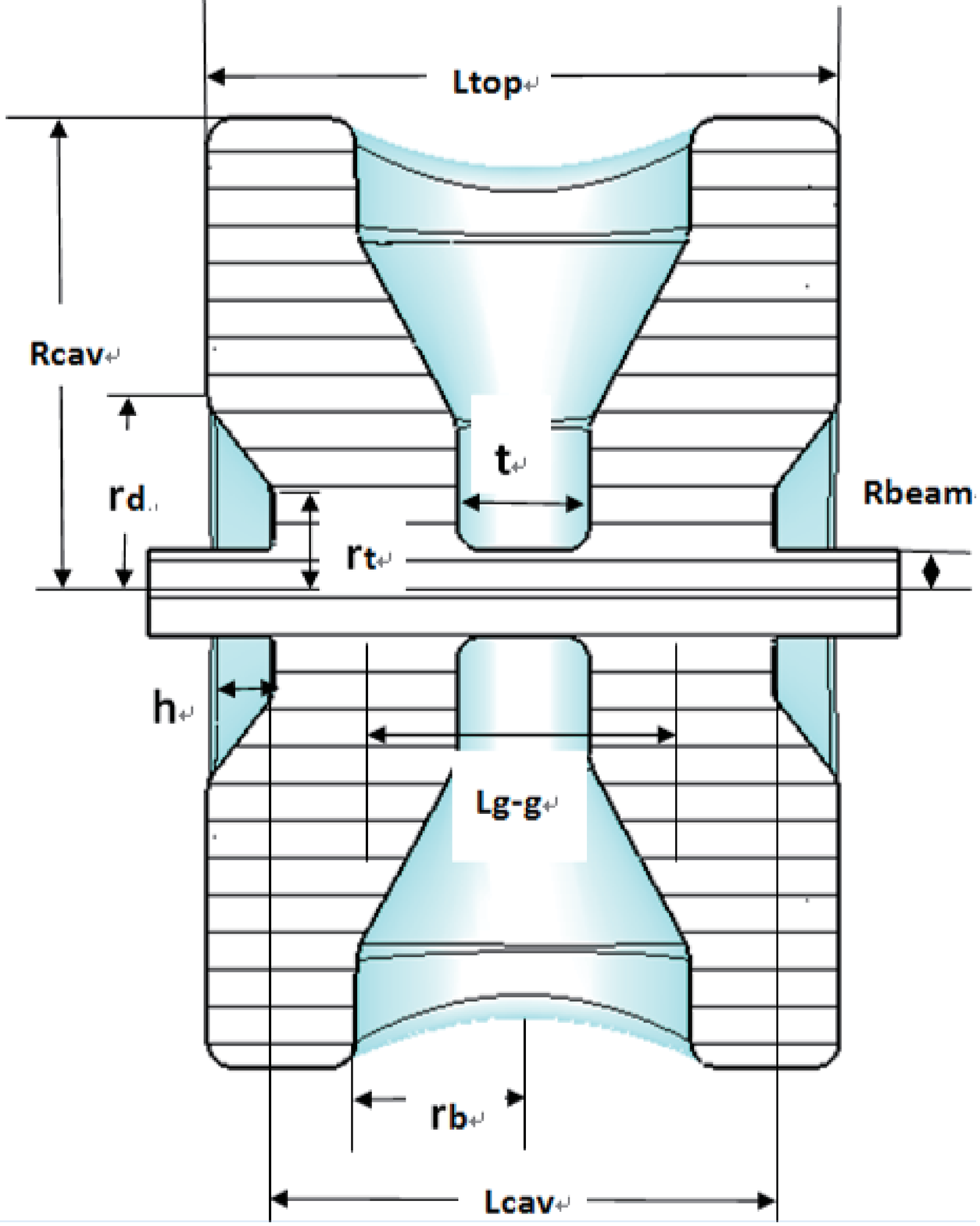}
\includegraphics[width=3.5cm]{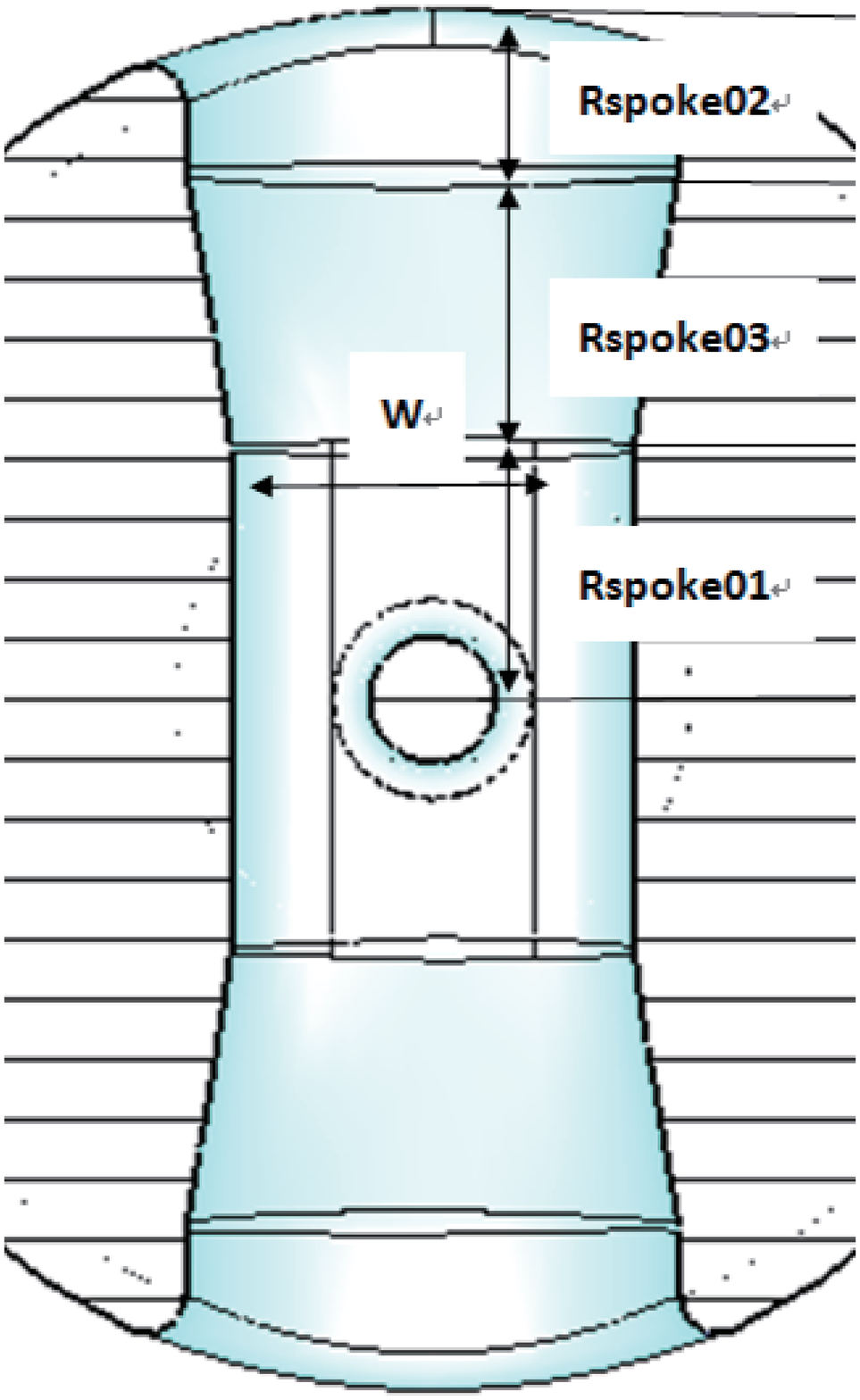}
\figcaption{\label{Fig1.}   Cut-away views of the spoke cavity model in the MWS and the main geometric parameters }
\end{center}

We used the original design progress to choose the main geometric parameters for the Spoke040 cavity \cite{lab4,lab5}.  First we chosen $Lg-g=\beta_g\lambda/2=185.0mm, Lcav=2/3*β_g\lambda=247.0mm, 2*rb/Lcav=0.4, W/Lcav=0.8 and Rspoke01=Rspoke02=Rspoke03$.  Then sweeping the main geometric parameters around those values to analysis how sensitive the RF parameters to the changes would be.  The results were showed in Fig2.

\begin{center}
\includegraphics[width=8.5cm]{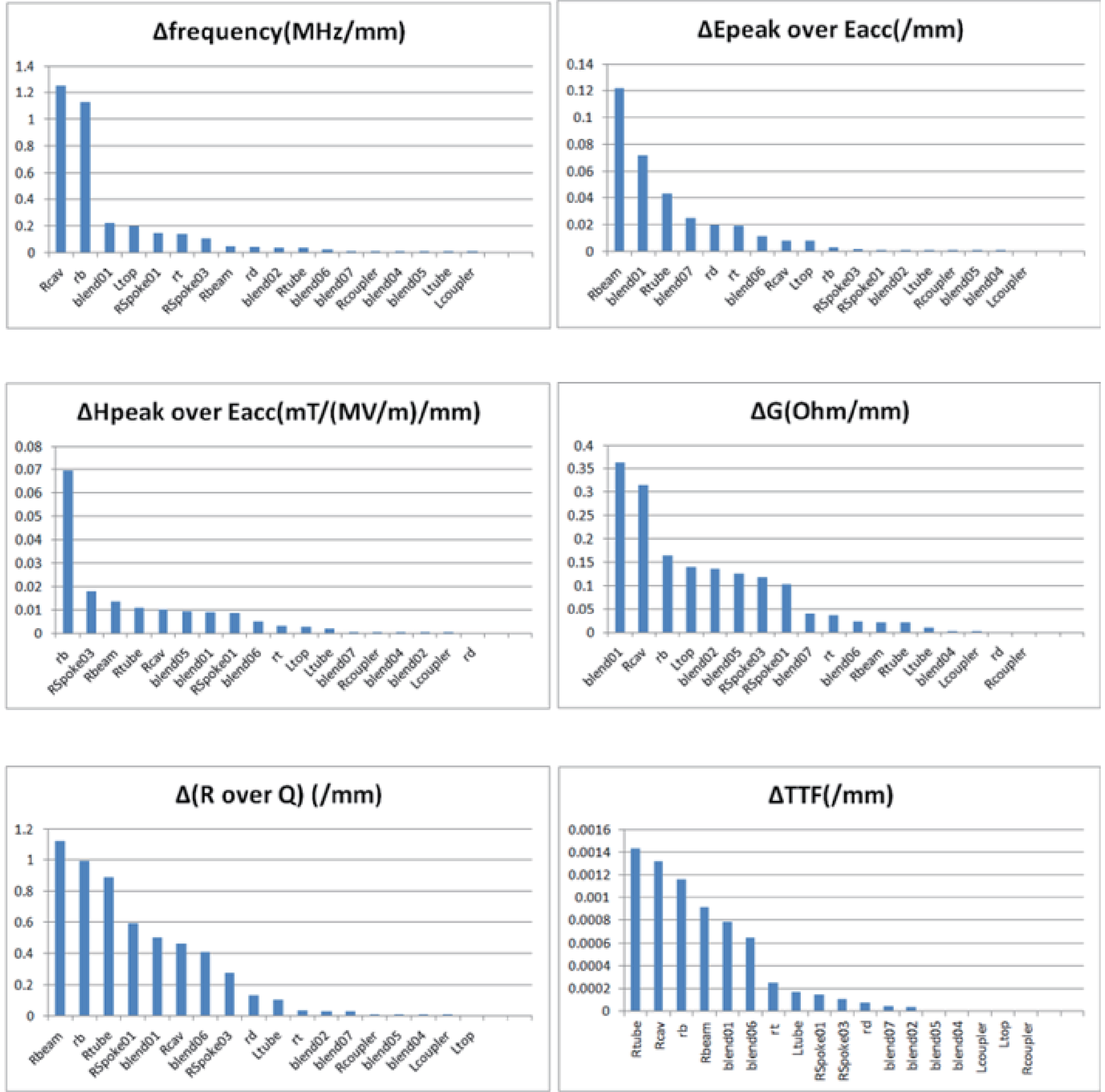}
\figcaption{\label{Fig2.}   The RF parameters change’ sensitive to geometric value }
\end{center}

Taking $r_d$(end wall bottom radius) as an example:when $\Delta r_d=1mm,  \Delta f=1.1MHz, \Delta($Epeak/Eacc$)=0.02, \Delta(Hpeak/Eacc)=(0.07mT/MV)/m, \Delta (R/Q)=1, \Delta G=0.15Ohm, \Delta TTF=0.001$.  The results showed the $r_d$ mostly influence $H_peak/E_acc, f and R/Q$; had less influence on TTF and G; and almost no influence on $E_peak/E_acc$.These provided us an effective criterion for more details’ RF design.

\subsection{RF optimization of the plate-shaped and arc-shaped end groups}

The Spoke040 cavity’s main sizes are much larger than that of Spoke021 and Spoke012 cavities. Considering improving the mechanical performance, we compared two types of end group to find a better geometric combination. Fig3 showed the difference between the end groups. Table1 showed the main geometric chose and table2 showed the RF parameters.

\begin{center}
\includegraphics[width=8.5cm]{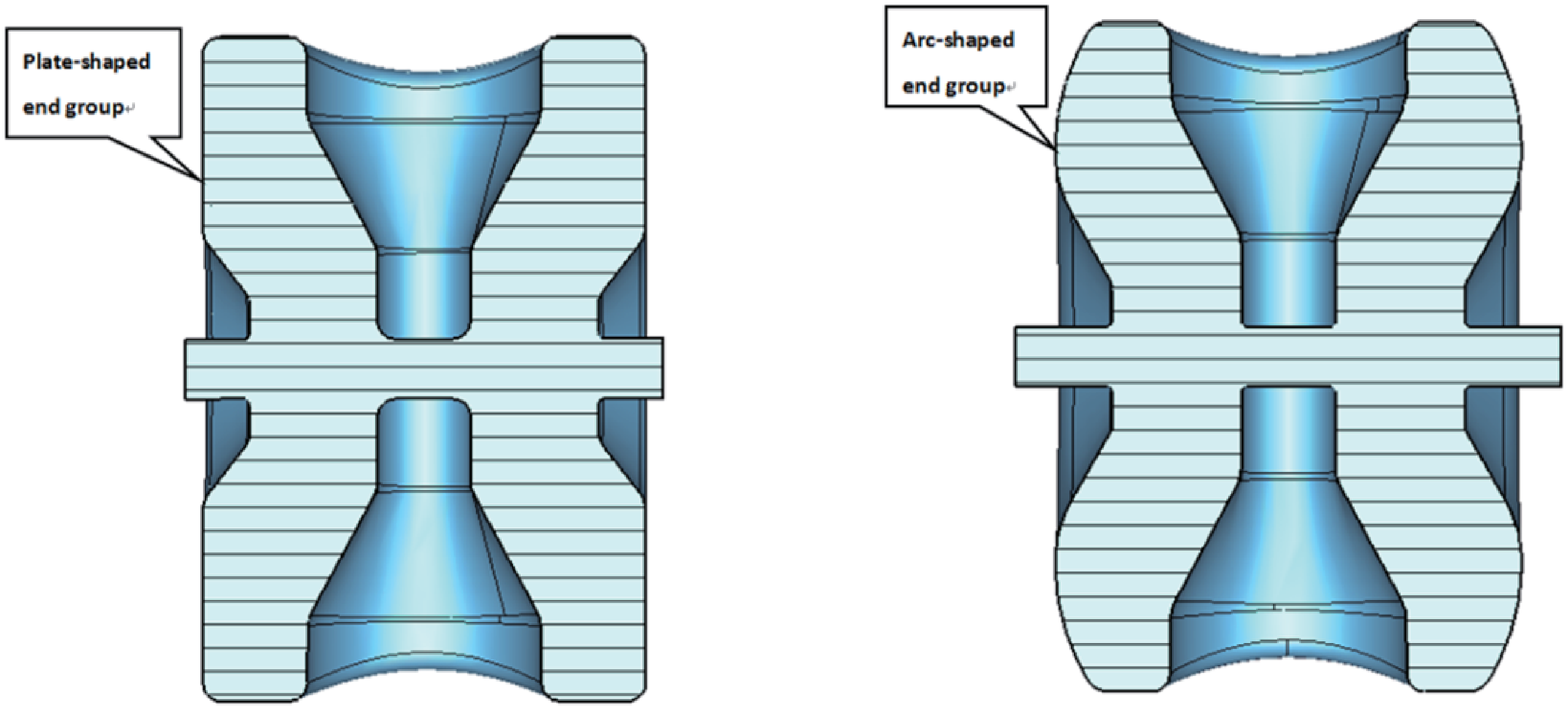}
\figcaption{\label{Fig3.}   Cut-away view of the plate-shaped (left) and arc-shaped (right) end groups }
\end{center}

\begin{center}
\tabcaption{ \label{Table1}  Geometric parameters of the two types end group of Spoke040}
\footnotesize
\begin{tabular*}{90mm}{@{\extracolsep{\fill}}ccccc}
\toprule  &$Ltop$/mm & $Lcav$/mm  & $Rcav$/mm &$W$/mm \\
\hline
Plate-shaped\hphantom{00} &370 &292 &278 &160 \\
Arc-shaped\hphantom{00} & 386.6 &292 & 278 &160\\
\bottomrule
\end{tabular*}%

\begin{tabular*}{90mm}{@{\extracolsep{\fill}}ccccc}
\toprule   &$Rbeam$/mm & $r_d$/mm  & $r_t$/mm  & $r_b$/mm\\
\hline
Plate-shaped\hphantom{00} &25 &110 &60 &98 \\
Arc-shaped\hphantom{00} &25 &120 &60 &98\\
\bottomrule
\end{tabular*}%
\end{center}

\begin{center}
\tabcaption{ \label{Table2}  Main RF parameters of the two types end group of Spoke040}
\footnotesize
\begin{tabular*}{90mm}{@{\extracolsep{\fill}}cccccc}
\toprule  $RF parameters$/ & Units  & $Plate-shaped$ &$Arc-shaped$ \\
\hline
f\hphantom{00} &MHz &324.41 &324.44  \\
$Epeak/Eacc$\hphantom{00} &M &2.82 &3.68 \\
$Hpeak/Eacc$\hphantom{00} &mT/(MV/m) &6.25 &8.31 \\
Q\hphantom{00} & &22923.0 &22795.6 \\
R/Q\hphantom{00} &Ω &247.22 &250.41 \\
TTF\hphantom{00} &  &0.817 &0.821 \\
\bottomrule
\end{tabular*}%
\end{center}

Compare to the plate-shaped end group, we just increased the Ltop and  $r_d$   of the arc-shaped end group to compensate the wave transverse volume to adjust the frequency on the basis of microwave perturbation theory. The main influence on RF parameters were the increases of  $Epeak/Eacc$ from 2.82 to 3.68 and $Hpeak/Eacc$ from 6.25 to 8.31 mT/(MV/m) .

\section{Mechanical studies}

We simulated the naked shell’s mechanical parameters in the room temperature $T=295.3K$ with niobium in different thickness and pressure while the beam ports were locked or not. The material properties of the niobium used are: Density $8600kg/m^3$, Young’s modulus $1.03e11Pa$, Poisson ratio 0.38. The results were shown in Fig4 and Table3.

\begin{center}
\includegraphics[width=4.0cm]{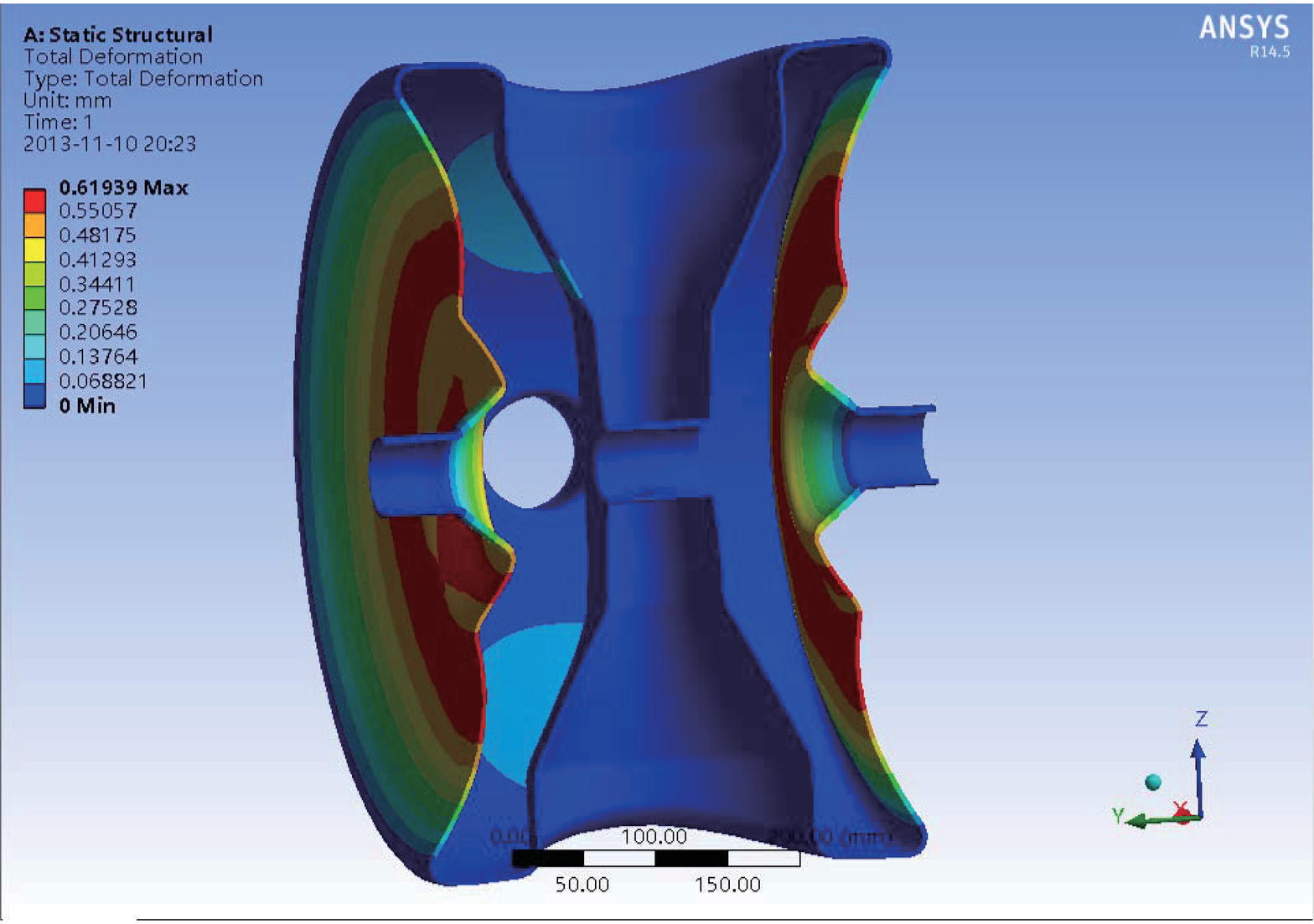}
\includegraphics[width=4.0cm]{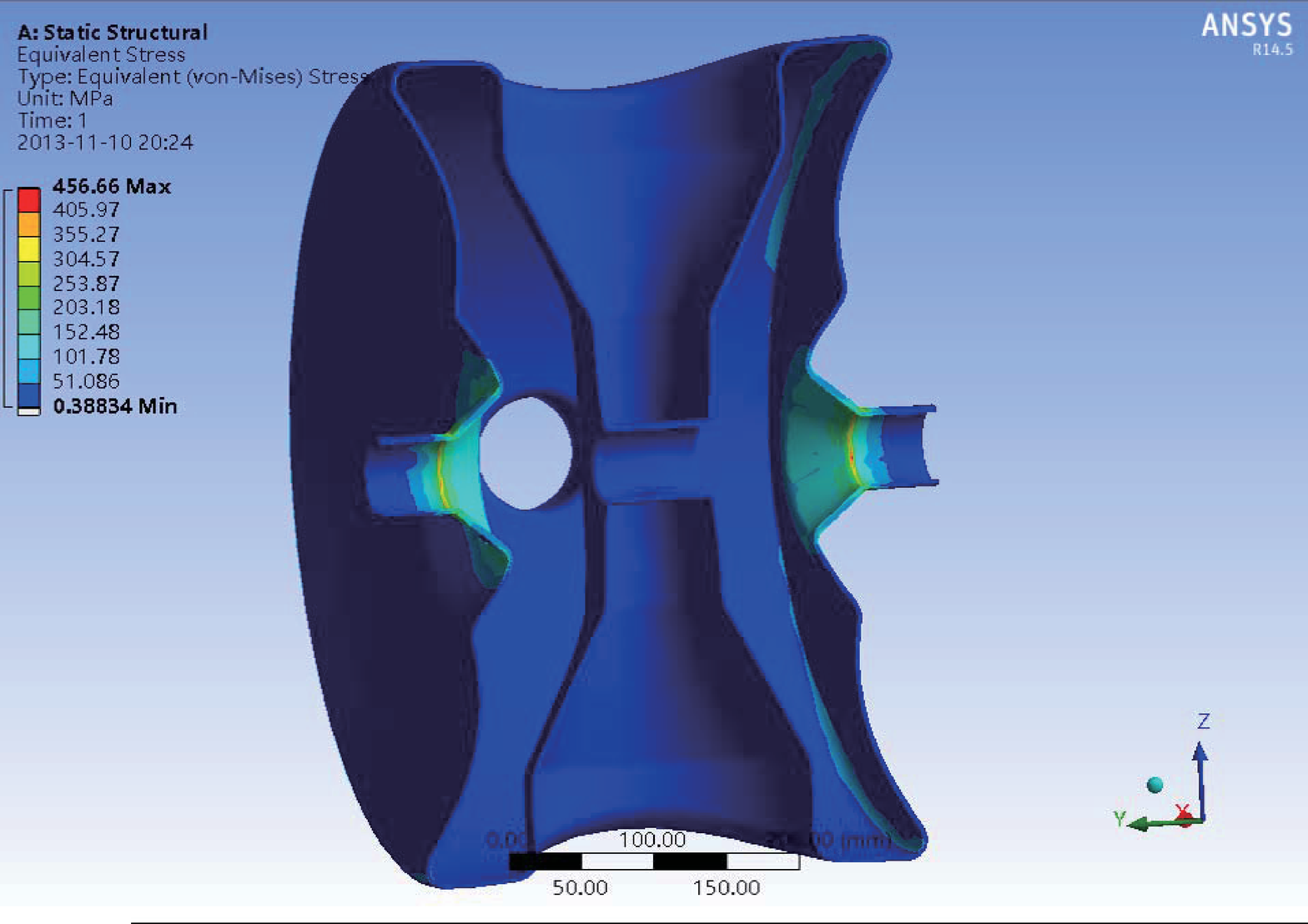}
\figcaption{\label{Fig4.}   Simulation of deformation (left) and stress (right) for Spoke040 @ the end group was plate-shaped, 3.0mm Nb shell, 1atm and free beam ports. }
\end{center}

\vspace{0.5cm}

\begin{center}
\tabcaption{ \label{Table3}  Some mechanical results for naked Spoke040 of different shapes, thickness, and pressure @ locked beam ports}
\footnotesize
\begin{tabular*}{80mm}{@{\extracolsep{\fill}}cccccccc}
\toprule  & Plate-shaped  \\
\toprule  Peak stress./MPa& 3.0mm &3.5mm  & 4.0mm  \\
\hline
1.0atm\hphantom{00} &556.56 &347.88 &448.58  \\
1.5atm\hphantom{00} &834.85 &521.81 &672.86  \\
2.0atm\hphantom{00} &1113.1 &695.75 &897.15  \\
\toprule  Deformation./mm& 3.0mm &3.5mm  & 4.0mm  \\
\hline
1.0atm\hphantom{00} &0.46617 &0.46611 &0.37654  \\
1.5atm\hphantom{00} &0.96476 &0.69916 &0.56481  \\
2.0atm\hphantom{00} &1.2863 &0.93221 &0.75308  \\
\bottomrule
\end{tabular*}

\begin{tabular*}{80mm}{@{\extracolsep{\fill}}cccccccc}
\toprule  & Arc-shaped  \\
\toprule  Peak stress./MPa& 3.0mm &3.5mm  & 4.0mm  \\
\hline
1.0atm\hphantom{00} &309 &362.2 &389.48  \\
1.5atm\hphantom{00} &463.5 &543.3 &584.22  \\
2.0atm\hphantom{00} &618 &724.39 &778.96  \\
\toprule  Deformation./mm& 3.0mm &3.5mm  & 4.0mm  \\
\hline
1.0atm\hphantom{00} &0.42168 &0.33315 &0.26586  \\
1.5atm\hphantom{00} &0.63251 &0.49972 &0.39879  \\
2.0atm\hphantom{00} &0.84335 &0.66629 &0.53172  \\
\bottomrule
\end{tabular*}%
\end{center}

The results shown the total deformation was mainly found symmetrically around the beam tube. And the max deformation appeared at the joint line of the arc and the slope, where the ring added should function to make the compromise. The press mainly gathered at the bottom of the beam tube symmetrically. The deformation and press decreased while the Nb shell’s thickness increased, but the material extensibility and hardness should also been concerned in the fabrication. Take the 3.0mm, 1atm situation as an example: the deformation and press of arc-shaped end group were much smaller than those of plate-shaped end group, which made us believe the sacrifice of the RF parameter in choosing the arc-shaped was worthy.

\section{Conclusion}

The study of the 325MHz, $\beta=0.40$ superconducting single spoke cavities was one of the key challenges for the China-ADS main linac, and also no precedent in the world. In this paper we introduced the optimization of the main geometric parameters, the studies of RF and naked mechanical parameters, and the compare of arc-shaped to plate-shaped end groups. The first SC Spoke040 cavity was just under construction. More works about the analysis of multipacting, the design of the rings and the study of the tuning sensitivity would be published. The work for the processing and the test were also in preparing.

\end{multicols}

\vspace{-1mm}
\centerline{\rule{80mm}{0.1pt}}
\vspace{2mm}

\begin{multicols}{2}

\end{multicols}

\clearpage

\end{document}